\documentclass[conference]{IEEEtran}
\IEEEoverridecommandlockouts
\usepackage{amsmath,amssymb}
\usepackage{graphicx}
\usepackage{cite}
\usepackage{tikz}\usetikzlibrary{decorations.pathreplacing,positioning}
\usepackage{wrapfig}
\usepackage[europeanresistors,americaninductors]{circuitikz}

\usepackage{color}
\usepackage{subcaption}
\usepackage{algorithm}
\usepackage{algpseudocode}

\makeatletter
\let\old@ps@headings\ps@headings
\let\old@ps@IEEEtitlepagestyle\ps@IEEEtitlepagestyle
\def\psccfooter#1{%
    \def\ps@headings{%
        \old@ps@headings%
        \def\@oddfoot{\strut\hfill#1\hfill\strut}%
        \def\@evenfoot{\strut\hfill#1\hfill\strut}%
    }%
    \def\ps@IEEEtitlepagestyle{%
        \old@ps@IEEEtitlepagestyle%
        \def\@oddfoot{\strut\hfill#1\hfill\strut}%
        \def\@evenfoot{\strut\hfill#1\hfill\strut}%
    }%
    \ps@headings%
}
\makeatother

\psccfooter{%
        \parbox{\textwidth}{\hrulefill \\ \small{24th Power Systems Computation Conference} \hfill \begin{minipage}{0.2\textwidth}\centering \vspace*{4pt} \includegraphics[scale=0.06]{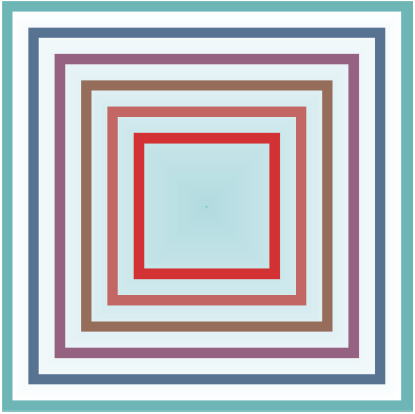}\\\small{PSCC 2026} \end{minipage} \hfill \small{Limassol, Cyprus --- June 8-12, 2026}}%
}

\title{Dynamic Passivity Multipliers for Plug-and-Play Stability Certificates of Converter-Dominated Grids}

\author{\IEEEauthorblockN{Andrey Gorbunov\IEEEauthorrefmark{1},
Youhong Chen\IEEEauthorrefmark{2},
Petr Vorobev\IEEEauthorrefmark{3}, 
Jin Ma\IEEEauthorrefmark{1},
Gregor Verbi\v{c}\IEEEauthorrefmark{1} }
\IEEEauthorblockA{\IEEEauthorrefmark{1} School of Electrical and Computer Engineering,
The University of Sydney,
Australia.}
\IEEEauthorblockA{\IEEEauthorrefmark{2} CAP Group,
Imperial College London,
UK.}
\IEEEauthorblockA{\IEEEauthorrefmark{3} School of EEE,
Nanyang Technological University,
Singapore}
}

\begin{document}
\maketitle
\begin{abstract}
Ensuring small-signal stability in power systems with a high share of inverter-based resources (IBRs) is hampered by two factors: (i) device and network parameters are often uncertain or completely unknown, and (ii) brute-force enumeration of all topologies is computationally intractable.  These challenges motivate plug-and-play (PnP) certificates that verify stability locally yet hold globally. Passivity is an attractive property because it guarantees stability under feedback and network interconnections; however, strict passivity rarely holds for practical controllers such as Grid Forming Inverters (GFMs) employing P-Q droop. This paper extends the passivity condition by constructing a dynamic, frequency-dependent multiplier that enables PnP stability certification of each component based solely on its admittance, without requiring any modification to the controller design. The multiplier is parameterised as a linear filter whose coefficients are tuned under a passivity goal. Numerical results for practical droop gains confirm the PnP rules, substantially enlarging the certified stability region while preserving the decentralised, model-agnostic nature of passivity-based PnP tests.
\end{abstract}


\section{Introduction}

The rapid growth of renewable energy integration has dramatically increased the complexity and uncertainty in power system dynamics. As the penetration of inverter\mbox{-}based resources (IBRs) continues to rise, ensuring small\mbox{-}signal stability becomes increasingly challenging due to fundamental changes in system behavior—including reduced inertia, faster dynamics, and more complex control interactions. Recent stability incidents in several major power systems illustrate these challenges: oscillations preceding the 2021 Iberian Peninsula blackout, Great Britain's sub\mbox{-} and super\mbox{-}synchronous oscillations, and similar events in the Australian grid \cite{AEMO_WMZ_2022, cheng2022real}. These incidents reveal that conventional stability assessment methods may no longer be adequate for converter\mbox{-}dominated grids.



Beyond the physical challenges of reduced system inertia, stability analysis of converter-dominated grids faces analytical and computational hurdles. Specifically, eigenvalue-based methods scale poorly with system size—characterizing the stability boundary becomes computationally prohibitive beyond even modest networks of 10 interconnected converters \cite{gorbunov2021identification}. As converter penetration increases, power system operators need scalable alternatives to conventional stability analysis. This context drives interest in compositional approaches with decentralised verification properties. Particularly promising are passivity-based methods, where stability guarantees compose naturally under interconnection \cite{sepulchre2012constructive}. The challenge, however, is that industry-standard control schemes like GFMs with P\,–\,Q droop exhibit frequency-dependent passivity violations, particularly in the low-frequency range where active power droops introduce negative damping effects \cite{dey2022passivity, vorobev2019decentralized}.

A range of decentralised remedies have therefore emerged, each with trade\mbox{-}offs. One line of work redesigns controllers (or adds filters/virtual impedances) to enforce passivity or bounded\mbox{-}real constraints \cite{watson2021scalable,pates2019robust,gross2022compensating,chen2024unified}, improving composability but often at the cost of either severe restrictions on control design or modifications that are too complex for practical deployment. Another strand retains decentralisation but relies on modelling simplifications—e.g., approximately uniform $R/X$ ratios, inductive (lossless) network assumptions, or neglecting internal voltage/current\mbox{-}control dynamics—which can limit applicability \cite{haberle2025decentralized,siahaan2024decentralized,huang2019plug,gorbunov2021identification,subotic2020lyapunov}. Impedance\mbox{-}based methods \cite{harnefors2007input} remain central but require careful choices between bus\mbox{-} and nodal\mbox{-}admittance formulations to preserve decentralisability \cite{jiang2023impedance}. Application\mbox{-}oriented studies—\cite{huang2024gain} among others—typically validate specific GFM designs under known network and controller parameters, while hybrid AC/DC investigations report low\mbox{-}frequency non\mbox{-}passivity that limits strict passivity tests \cite{dey2022passivity,watson2020control}. Closer to our aims, \cite{vorobev2019decentralized} proposed premultiplying each device admittance $Y(j\omega)$ by a frequency\mbox{-}dependent multiplier $m(\omega)$ so that $m(\omega)Y(j\omega)$ is \textit{positive real} (i.e., Hermitian part is positive definite) preserving decentralised composition without enforcing strict passivity on the raw device; however, that work did not provide a systematic procedure to synthesise such multipliers. Our generalised passivity approach retains decentralisation and plug\mbox{-}and\mbox{-}play composability while overcoming these limitations by (i) removing the need for detailed device/network models or controller redesign, (ii) avoiding lossless or uniform~\mbox{$R/X$} assumptions, and (iii) scaling naturally to MIMO dynamics rather than SISO frequency loops as in \cite{pates2019robust,gross2022compensating}.


Contributions are summarised as follows:
\begin{itemize}
  \item We propose a systematic method for discovering plug-and-play stability certificates through dynamic multiplier synthesis that: 
  \begin{itemize}
  \item makes no assumptions about power system components models, accommodating arbitrary internal control loops without simplifications; 
  \item imposes no restrictions on network characteristics such as lossless network, static behaviour, or homogeneity, fully accounting for electromagnetic network dynamics; and 
  \item can potentially be generalised to a black-box approach without requiring the knowledge of explicit components models.
  \end{itemize}
  \item We validate the approach on the IEEE 39-bus system using detailed small-signal models, confirming the plug-and-play stability.
\end{itemize}

The rest of this paper is structured as follows. Section \ref{sec:Preliminaries} provides a brief overview of the dynamic multiplier framework introduced in \cite{vorobev2019decentralized}, establishing the theoretical foundation for our work. Section \ref{sec:multipliers-synthesis} describes our systematic approach to dynamic multiplier synthesis, where we parametrise multipliers by state-space filters and tune them using optimisation techniques. In Section \ref{sec:CaseStudies}, we present numerical studies that validate our approach, including tests on a two-bus system and the IEEE 39-bus system with randomly allocated GFM converters. Finally, Section \ref{sec:Conclusion} summarises our contributions and outlines directions for future research.

\section{Preliminaries}\label{sec:Preliminaries}
\begin{figure}
  \centering
  \includegraphics[width=0.6\columnwidth]{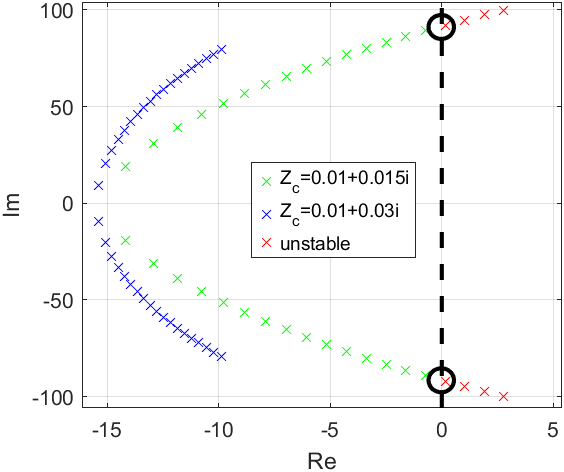}
  \caption{Trajectory of the eigenvalues for two systems each consisting of two GFM inverters during a homotopy. Eigenvalues of one of the system move into the right-half plane (unstable region). Eigenvalues that cross the imaginary axis are marked with “$\circ$".}
  \label{fig:two-inv-root-locus}
\end{figure}

This section provides a brief description of the dynamic multiplier framework that lies in the foundation of the present work. The initial theoretical concept was first introduced in \cite{vorobev2019decentralized}. The main idea of the method is to perform a \textit{homotopy} - a gradual transform - of the system from a certain (fictitious) definitely stable configuration to the configuration of interest. Then, stability of the system of interest can be certified if we ensure that none of the system eigenvalues crossed the imaginary axis as the homotopy is performed - if all the system eigenvalues remain in the left–half plane throughout homotopy, the final
system is certified stable.  The example of such a homotopy for two different systems, each composed of GFM inverters interconnected through impedance $Z_c$, is illustrated in Fig. \ref{fig:two-inv-root-locus}. The first system is tuned to be stable, and its eigenvalues move within the left half-plane during the homotopy. The second system, on the contrary, loses stability, and this inevitably leads to some (at least one) of the eigenvalues crossing the imaginary axis during the homotopy.




As was demonstrated in \cite{vorobev2019decentralized}, one can ensure that none of the system eigenvalues crossed the imaginary axis by demanding the following condition to be satisfied for every power system component (including lines):
\begin{equation}
S_i(j\omega)=\operatorname{Her}\!\big(m(\omega)Y_i(j\omega)\big) \succ 0,
\label{eq:Sdef}
\end{equation}
\noindent
here $\operatorname{Her}(X):=\tfrac12\big(X+X^\dagger\big)$ denotes the Hermitian part of a matrix. Matrices \(Y_i(j\omega)\) are $2$ by $2$ admittance matrices of the power system components in a $dq$ reference frame. For more details on the dynamic impedance/admittance representation of power system components the reader is referred to the classical work \cite{harnefors2007input}. Matrix   
\(m(\omega)\) (which is also $2$ by $2$) is non-singular, but does not have to be sign (semi)definite, and can be a function of frequency, including a piece-wise constant function. A suitably chosen \(m(\omega)\) can make the admittances of all the components ``look'' passive; during the homotopic transformation, no eigenvalues cross the imaginary axis, ensuring stability without the conservatism of classical passivity approach. The following subsection provides a justification for the equation \eqref{eq:Sdef}.

\subsection{Network / device decomposition}

Consider an \(n\)-bus reduced network in synchronous \(dq\) coordinates. Let
\(v=\operatorname{col}(v_1,\dots,v_n)\), \(i_{inj}=\operatorname{col}(i_1,\dots,i_n)\),
where \(v_k,i_k\in\mathbb{C}^2\) are the local phasor (or small-signal) voltage and injected currents at bus \(k\).
The total nodal admittance splits as
\begin{equation}
i_{inj} = Y_{\text{tot}}(s) v,\qquad
Y_{\text{tot}}(s)=Y_{\text{net}}(s)+Y_D(s),
\end{equation}
where $Y_D(s)=\operatorname{blkdiag}\big(Y_1(s),\dots,Y_n(s)\big)$
collects the (converter / load) device admittances $Y_k(s)$, and \(Y_{\text{net}}(s)\) is the usual sparse nodal admittance matrix assembled from the series line (or \(\pi\)-) models. For passive RL (or RL+shunt) lines each off-diagonal entry associated with line \(\ell=(k,m)\) is \(-Y_{\ell}(s)\), and each diagonal entry is the sum of incident line admittances. Hence
\begin{equation}
Y_{\text{net},km}(s)=
\begin{cases}
-\;Y_{\ell}(s), & k\neq m,\\
\sum_{m\in\mathcal{N}_k} Y_{(k,m)}(s), & k=m,\\[2pt]
0,& \text{otherwise.}
\end{cases}
\end{equation}

In this work, we consider the network to consist of RL lines. Each line admittance in the $dq$-frame is given by (see the derivations in \cite{vorobev2019decentralized}):
\begin{equation}
Y_{\ell}(s) = Z_{\ell}^{-1}(s) = 
\begin{bmatrix}
R_{\ell} + sL_{\ell} & -\omega_0 L_{\ell} \\
\omega_0 L_{\ell} & R_{\ell} + sL_{\ell}
\end{bmatrix}^{-1}
\end{equation}
where $R_{\ell}$ and $L_{\ell}$ are the resistance and inductance of line $\ell$, and $\omega_0$ is the nominal system frequency.
For device admittances $Y_k$, this work focuses on droop-controlled GFM converters. Due to the complexity of these models with multiple cascaded control loops, we do not provide closed-form expressions for their admittances as they would be unnecessarily bulky. The complete details of the GFM converter model, including control structure and parameters, are given in Section \ref{sec:CaseStudies}.

For the homotopic argument, we introduce a scalar parameter \(\alpha\in[0,1]\) and a continuously varying family.
\noindent Homotopies (continuous in $\alpha\in[0,1]$):
\begin{align}
Y_{\text{tot}}(s,\alpha)&=Y_{\text{net}}(s,\alpha)+Y_D(s,\alpha),\\
Y_{\text{tot}}(s,0)&=Y_{\text{ref}}(s),\qquad
Y_{\text{tot}}(s,1)=Y_{\text{tot}}(s).
\end{align}
All derivations below apply pointwise in \(\alpha\); we make explicit that the required inequalities must hold \emph{uniformly} for every \(\alpha\in[0,1]\).

In order to come to \eqref{eq:Sdef} we first note, that in the absence of the externally injected nodal currents the system dynamics must satisfy the following set of conditions: 
\begin{equation}
\big[Y_{\text{net}}(s)+Y_D(s)\big]v = 0
\end{equation}
This system has non-trivial solutions only for certain values of $s=s_k$ (which coincide with the system eigenvalues) that guarantee:
\begin{equation}
\textrm{Det}\big[Y_{\text{net}}(s)+Y_D(s)\big] = 0
\end{equation}
Thus, the condition that none of the system eigenvalues cross the imaginary axis reduces to the following:   
\begin{equation}\label{det_omega}
\textrm{Det}\big[Y_{\text{net}}(j\omega,\alpha)+Y_D(j\omega,\alpha)\big] \neq 0
\end{equation}
This condition must hold all the way through the homotopy, i.e. for  \(\alpha\in[0,1]\). On the other hand, condition \eqref{det_omega} will hold if we multiply it's left-hand by any nonsingular matrix \(M(\omega)\):    
\begin{equation}\label{det_M_omega}
\textrm{Det}\big[M(\omega)Y_{\text{net}}(j\omega,\alpha)+M(\omega)Y_D(j\omega,\alpha)\big] \neq 0
\end{equation}
Finally, instead of conditioning on the matrix determinant, we will demand the positive definiteness (PD) of its Hermitian part:
\begin{equation}\label{Y_pos}
\operatorname{Her}\!\big[M(\omega)Y_{\text{net}}(j\omega,\alpha)+M(\omega)Y_D(j\omega,\alpha)\big] \succ 0 
\end{equation}
The latter step can introduce some conservativeness, but allows us to formulate the fully decentralised conditions \eqref{eq:Sdef} by exploiting the structure of the network admittance matrix, as outlined in the following subsection. For full details of the derivation, see the original paper \cite{vorobev2019decentralized}.

\subsection{Uniform multiplier and simplified decentralised condition}

Assume a \emph{uniform} dynamic multiplier
\begin{equation}
M(s)= \operatorname{blkdiag}\big(m(s), \dots, m(s) \big),
\end{equation}
where $ m(s)\in\mathbb{C}^{2\times 2}\ \text{nonsingular}$.
Then \(S(j\omega,\alpha)\) in (\ref{eq:Sdef}) separates into device and network parts:
\begin{equation}
S(j\omega,\alpha)= S_D(j\omega,\alpha)+S_{\text{net}}(j\omega,\alpha),
\end{equation}
with $S_D = \operatorname{blkdiag}\big(S_1,\dots,S_n\big), \ S_k(j\omega,\alpha) = \operatorname{Her}\big(m(j\omega)Y_k(j\omega,\alpha)\big)$.

The network contribution can be written as a sum of per–line matrices:
\begin{equation}
S_{\text{net}}(j\omega,\alpha)=\sum_{\ell=(i,j)\in\mathcal{E}} \widetilde S_\ell(j\omega,\alpha),
\end{equation}
where for each line \(\ell=(i,j)\), we define a 4×4 matrix
\[
B_\ell(j\omega,\alpha)=
\begin{bmatrix}
m(j\omega)Y_{\ell}(j\omega,\alpha) & -\,m(j\omega)Y_{\ell}(j\omega,\alpha)\\[4pt]
-\,m(j\omega)Y_{\ell}(j\omega,\alpha) & \;\;m(j\omega)Y_{\ell}(j\omega,\alpha)
\end{bmatrix},
\]
and its Hermitian part \(\hat S_\ell(j\omega,\alpha):=\operatorname{Her}\big(B_\ell(j\omega,\alpha)\big)\).

The matrix \(\widetilde S_\ell(j\omega,\alpha)\) is obtained by embedding \(\hat S_\ell\) into the full system matrix via a suitable permutation matrix that maps the line's terminal buses to their positions in the network. Importantly, the non-zero eigenvalues of \(\widetilde S_\ell\) are identical to those of \(\hat S_\ell\), which in turn are determined by the eigenvalues (up to the scalar factor 2) of \(m(j\omega)Y_\ell(j\omega,\alpha)\).

Therefore, for each line \(\ell=(i,j)\), we have the equivalence:
\[
\widetilde S_\ell(j\omega,\alpha)\succ 0
\Longleftrightarrow
S_\ell(j\omega,\alpha) :=\operatorname{Her}\big(m(j\omega)Y_\ell(j\omega,\alpha)\big)\succ 0.
\]

Summing all line contributions with the diagonal device terms gives
\[
S(j\omega,\alpha)= \operatorname{blkdiag}(S_1,\dots,S_n)\;+\;\sum_{\ell=(i,j)}
\widetilde S_\ell(j\omega,\alpha).
\]
Therefore, a frequency– and homotopy–wise sufficient (and decentralised) condition for
\(S(j\omega,\alpha)\succ 0\) for every \(\alpha\in[0,1]\) is
\begin{equation}
\label{eq:decentralised-conditions}
\begin{aligned}
S_k(j\omega, \alpha) &\succ 0, \forall k,\ \forall \omega > 0,\ \forall \alpha\in[0,1],\\[3pt]
S_\ell(j\omega, \alpha) &\succ 0, \forall \ell\in\mathcal{E},\ \forall \omega > 0,\ \forall \alpha\in[0,1].
\end{aligned}
\end{equation}

\paragraph*{Decentralised verification}
Each device (bus) verifies \(S_k\succ 0\) over \(\omega\) and \(\alpha\) using only its own \(Y_k(\cdot,\alpha)\) and the common \(m\).
Each line (or either terminal device) verifies \(\operatorname{Her}(m Y_{ij})\succ 0\) over \(\omega,\alpha\).

Thus, under a uniform multiplier, certifying converter (device) and line blocks individually for all \(\alpha\in[0,1]\) and \(\omega > 0\) is sufficient for global small-signal stability. The next section presents the main result of this paper—a systematic search procedure for dynamic multipliers that satisfy the decentralised condition \eqref{eq:decentralised-conditions}.

\section{Dynamic Multipliers Synthesis}\label{sec:multipliers-synthesis}
\subsection{Homotopy-free certificates}
Firstly, we further simplify certificates in \eqref{eq:decentralised-conditions} by eliminating the need to check conditions over the entire homotopy path \(\alpha\in[0,1]\).
We first formalise decentralised stability conditions in \eqref{eq:decentralised-conditions} as
\begin{equation}
\mathcal{C}(\alpha):\;
\begin{cases}
S_k(j\omega, \alpha) \succ 0,& \forall k,\ \forall \omega>0,\\
S_\ell(j\omega, \alpha) \succ 0,& \forall \ell\in\mathcal{E},\ \forall \omega>0.
\end{cases}
\label{eq:C1}
\end{equation}

\noindent\textbf{Lemma (Endpoint equivalence under linear homotopy).}
Suppose the multiplier \(m(s)\) is independent of \(\alpha\) and each device and line admittance is linear in \(\alpha\):
\begin{equation}
\begin{aligned}
Y_k(s,\alpha) &= (1-\alpha)Y_{k,\text{ref}}(s)+\alpha Y_k(s),\\
Y_{ij}(s,\alpha) &= (1-\alpha)Y_{ij,\text{ref}}(s)+\alpha Y_{ij}(s)
\end{aligned}
\label{eq:affineY}
\end{equation}
for \(\alpha\in[0,1]\). Then
\[
\mathcal{C}(\alpha)\ \text{holds for every }\alpha\in[0,1]
\quad\Longleftrightarrow\quad
\mathcal{C}(0)\ \text{and}\ \mathcal{C}(1)\ \text{hold}.
\]

\emph{Proof.}
(\(\Rightarrow\)) If all inequalities in \eqref{eq:C1} hold for every \(\alpha\), they hold at \(\alpha=0,1\).
(\(\Leftarrow\)) For each bus \(k\) and frequency \(\omega\),
\[
\operatorname{Her}\!\big(mY_k(\alpha)\big)=
(1-\alpha)\operatorname{Her}\!\big(mY_{k,\text{ref}}\big)+\alpha \operatorname{Her}\!\big(mY_{k}\big),
\]
a linear function of \(\alpha\). The set of PD matrices is convex; thus, PD at \(\alpha=0\) and \(\alpha=1\) implies PD for all \(\alpha\in[0,1]\). The same argument applies to every line admittance \(Y_{ij}\). Hence \eqref{eq:C1} holds for all \(\alpha\). \hfill$\square$

\noindent\textbf{Remark (No benefit from non-linear homotopies).}
The linear homotopy approach is optimal for our passivity-based certificates. If either endpoint violates the conditions in \eqref{eq:C1}, no other homotopy path—regardless of complexity—can fix it. Conversely, if both endpoints satisfy the conditions, the linear path is already sufficient due to convexity. Therefore, using more complex non-linear homotopies provides no benefit for our passivity-based certificates. Note that this applies specifically to our sufficient conditions; for necessary-and-sufficient stability criteria based on determinants, non-linear paths might still offer advantages.

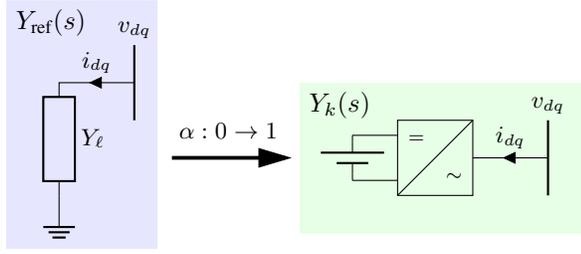
\begin{figure}
  \centering
  \begin{tikzpicture}[american voltages]
    \fill[blue!10] (-1.2,-0.2) rectangle (0.8,3.1);
    \node[anchor=north west] at (-1.2,3.1) {$Y_{\text{ref}}(s)$};
    
    \draw[thick] (0.5,2.5) -- (0.5,1.5);
    \node at (0.5,2.7) [font=\small]{$v_{dq}$};
    
    \draw (0.5,2.0)
      to[short, i>_=$i_{dq}$, font=\small] (-0.5,2.0)
        to[R, l=$Y_\ell$, font=\small] (-0.5,0.5) node[ground]{};

    \draw[line width=0.5mm, -{Triangle[length=4.5mm,width=2.5mm]}] (1.0,1) -- (2.6,1);
    \node[above=4pt, font=\small] at (1.75,1) {$\alpha: 0\to 1$};


    \fill[green!10] (2.7,0.0) rectangle (6.5,2.0);
    \node[anchor=north west] at (2.7,2.0) {$Y_k(s)$};
    
    \draw (4.0,0.5) rectangle (5.0,1.5);
    
    \draw (3.4,1.3) -- (4.0,1.3);
    \draw (3.4,0.7) -- (4.0,0.7);
    
    \draw (3.4,1.3) to[battery1] (3.4,0.7);

    \node at (4.25,1.25) {\scriptsize $=$};   
    \node at (4.75, 0.75) {\scriptsize $\sim$}; 
    
    \draw (4.0,0.5) -- (5.0,1.5); 
    
    \draw (6.0,1) to[short, i>_=$i_{dq}$,font=\small] (5.0,1);
    
    \draw[thick] (6.0,1.5) -- (6.0,0.5);
    \node at (6.0,1.7) [font=\small]{$v_{dq}$};
  \end{tikzpicture}
  \caption{Homotopic transformation from a passive reference system ($\alpha=0$) to the actual converter-based system ($\alpha=1$). With passive line admittances as references, only the endpoint at $\alpha=1$ needs verification for stability certification.}
  \label{fig:homotopy}
\end{figure}

This approach can be further simplified by choosing passive line admittances as references for device admittances. Since transmission lines are inherently passive, a network composed entirely of passive elements is guaranteed stable, as illustrated in Fig. \ref{fig:homotopy}. With this reference choice, we need only verify the conditions at $\alpha=1$ since the conditions at $\alpha=0$ are already satisfied. Thus, our stability certification simplifies to:

\begin{equation}
\label{eq:decentralised-conditions-1}
\boxed{
\begin{aligned}
\operatorname{Her}\big(m(j\omega)Y_k(j\omega)\big) &\succ 0, \forall k,\ \forall \omega > 0,\\[3pt]
\operatorname{Her}\big(m(j\omega)Y_\ell(j\omega)\big) &\succ 0, \forall \ell\in\mathcal{E},\ \forall \omega > 0.
\end{aligned}
}
\end{equation}

\subsection{Parametric Multiplier Design}

Having simplified the stability certification to checking only the endpoint conditions, we now address how to systematically design effective multipliers. In the original work \cite{vorobev2019decentralized}, the multiplier \(m(j\omega)\) took a specific piecewise-constant form:
\begin{equation} \label{eq:piecewise}
m(j\omega) = 
\begin{cases}
\begin{bmatrix} 0 & -1 \\ 1 & 0 \end{bmatrix}, & \omega < \omega_f \\
I_2, & \omega \geq \omega_f
\end{cases}
\end{equation} 
where $\omega_f$ is an implicit parameter (e.g., $\omega_f = 2\pi \times 50 \,\text{rad/s}$). This multiplier applies a $90^\circ$ rotation in the complex plane at low frequencies, effectively compensating for the non-passive behavior typical of droop controllers in that range. At higher frequencies, where passivity is naturally satisfied, the multiplier reverts to the identity. Although this piecewise-constant structure is conceptually simple, it tends to be overly conservative—restricting the certified stability region.

\begin{figure}
  \centering
  \begin{tikzpicture}[>=latex,
      node distance=2.0cm,
      block/.style={draw, thick, rectangle, minimum height=0.8cm,
                    minimum width=1.5cm, align=center,font=\small}]
    
    \node[coordinate] (in) {};
    \node[right=0.5cm of in] (split) {};
    
    \node[block, fill=green!10, above right=1.0cm and 1.2cm of split] (Y1) {$Y_1(s)$};
    \node[block, right=1.0cm of Y1] (M1) {$m(s)$};
    
    \node[block, fill=green!10, right=1.0cm and 1.2cm of split] (Y2) {$Y_2(s)$};
    \node[block, right=1.0cm of Y2] (M2) {$m(s)$};
    
    \node at (4.0,-1.5) (dots) {$\vdots$};
    
    \node[block, fill=green!10, below right=2.0cm and 1.2cm of split] (Yn) {$Y_n(s)$};
    \node[block, right=1.0cm of Yn] (Mn) {$m(s)$};
    
    \node[block, fill=blue!10, below right=3.7cm and 1.2cm of split] (Yl) {$Y_{ij}(s)$};   \node[block, right=1.0cm of Yl] (Ml) {$m(s)$};
    
    \node[coordinate, right=1.0cm of M1] (out1) {};
    \node[coordinate, right=1.0cm of M2] (out2) {};
    \node[coordinate, right=1.0cm of Mn] (outn) {};
    \node[coordinate, right=1.0cm of Ml] (outl) {};
    
    \draw[->] (in) |- node[above, near end, font=\scriptsize] {$v$} (Y1.west);
    \draw[->] (in) -- node[above, font=\scriptsize] {$v$} (Y2.west);
    \draw[->] (in) |- node[above, near end, font=\scriptsize] {$v$} (Yn.west);
    \draw[->] (in) |- node[above, near end, font=\scriptsize] {$v$} (Yl.west);
    
    \draw[->] (Y1) -- node[above, font=\scriptsize] {$i_1$} (M1);
    \draw[->] (Y2) -- node[above, font=\scriptsize] {$i_2$} (M2);
    \draw[->] (Yn) -- node[above, font=\scriptsize] {$i_n$} (Mn);
    \draw[->] (Yl) -- node[above, font=\scriptsize] {$i_{ij}$} (Ml);
    
    \draw[->] (M1) -- node[above, font=\scriptsize] {$y_1$} (out1);
    \draw[->] (M2) -- node[above, font=\scriptsize] {$y_2$} (out2);
    \draw[->] (Mn) -- node[above, font=\scriptsize] {$y_n$} (outn);
    \draw[->] (Ml) -- node[above, font=\scriptsize] {$y_{ij}$} (outl);
    
    \draw[decoration={brace, mirror, raise=5pt}, decorate] 
      (Y1.south west) -- (M1.south east) 
      node[midway, below=10pt, font=\scriptsize] {Passive};
    
    \draw[decoration={brace, mirror, raise=5pt}, decorate] 
      (Y2.south west) -- (M2.south east) 
      node[midway, below=10pt, font=\scriptsize] {Passive};
    
    \draw[decoration={brace, mirror, raise=5pt}, decorate] 
      (Yn.south west) -- (Mn.south east) 
      node[midway, below=10pt, font=\scriptsize] {Passive};
    
    \draw[decoration={brace, mirror, raise=5pt}, decorate] 
      (Yl.south west) -- (Ml.south east) 
      node[midway, below=10pt, font=\scriptsize] {Passive};
      
    \node[above=0.2cm of Y1, font=\small] {Tuning same $m(s)$ for all components};
    
  \end{tikzpicture}
  \caption{Decentralised stability certification: a single multiplier $m(s)$ is tuned to make all device admittances $Y_k(s)$ and line admittances $Y_{ij}(s)$ passive when combined in series. Each component is certified independently with the same multiplier.}
  \label{fig:passive_series}
\end{figure}
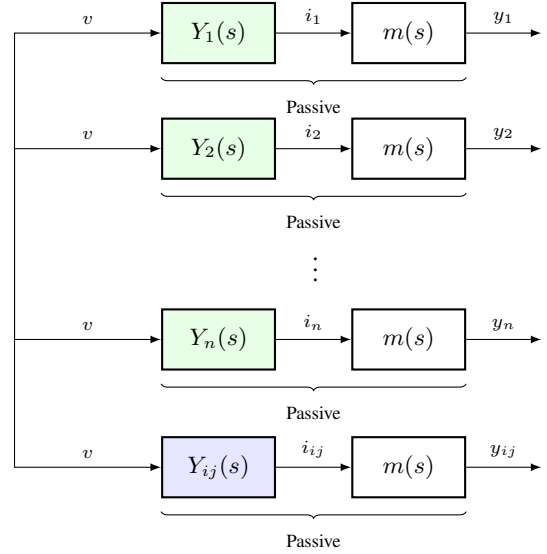

We extend (\ref{eq:piecewise}) by \emph{parametrising} \(m(j\omega)\) as a state-space filter of fixed order $m_f$:
\begin{equation}
m(s)=C_m(sI-A_m)^{-1}B_m+D_m,
\end{equation}
and tuning the coefficients of series interconnection \(m(s)Y(s)\) as illustrated in Fig. \ref{fig:passive_series} with \textsc{MATLAB}'s \texttt{systune} under a \texttt{PassivityGoal}. Because \(M(s)\) serves only as an analytic
certificate, no changes to the inverter firmware are required.

\subsection{Controller Synthesis with PassivityGoal}
The synthesis procedure operates according to Algorithm 1.
\begin{algorithm}[t]
\caption{Dynamic Multiplier Synthesis}
\begin{algorithmic}[1]
\State Initialise a state-space model for $m(s)$ with fixed order $m_f$, having free parameters in matrices $A_m$, $B_m$, $C_m$, and fixed $D_m = I_2$
\State Form the series interconnection $G_k(s) = m(s)Y_k(s)$ for each device and line admittance $Y_k(s)$
\State Define a \texttt{PassivityGoal} for each $G_k(s)$ requiring:
\[
  G_k(j\omega) + G_k^H(j\omega) \succeq 0 \quad \forall \omega > 0
\]
\State Execute \texttt{systune} optimisation to find parameters enforcing passivity across all $G_k(s)$
\end{algorithmic}
\end{algorithm}

In MATLAB's \texttt{systune}, passivity is enforced via the sector/scattering framework. Given a set of admittances $\{Y_k(s)\}$ and a fixed-order filter $m(s;\theta)$, define
\begin{equation}
R_k(s;\theta)\;=\;\big(I - m(s;\theta)\,Y_k(s)\big)\,\big(I + m(s;\theta)\,Y_k(s)\big)^{-1} ,
\end{equation}
where $\theta$ are the free parameters in $m(s;\theta)$, i.e. the entries of $A_m$, $B_m$, $C_m$.
Provided $I + mY_k$ is minimum-phase, passivity is equivalent to $\|R_k\|_{\infty} := \operatorname{max}_\omega \sigma(R_k(j\omega))\le 1$ for all $k$ \cite{schaft1996l2}.
The synthesis problem is posed as the multi-model nonsmooth minimax
\begin{equation}
\min_{\theta}\ \max_{k}\ 
\left\|\,R_k(\theta)\,\right\|_{\infty}
\qquad\text{(target }\le 1\text{)} .
\label{eq:minimax}
\end{equation} 
This objective is nonsmooth and nonconvex (maximising over models, frequencies, and singular values), so \texttt{systune} employs the Apkarian--Noll optimisation framework: computing exact $\mathcal{H}_\infty$ peaks, extracting subgradients from active singular vectors, and applying specialised descent/trust-region techniques with multi-start initialisation to mitigate local minima \cite{apkarian2006nonsmooth}.

While Algorithm 1 uses device admittances $Y_k(s)$ given as rational transfer functions, our approach doesn't assume any special internal structure of the model. Moreover, the algorithm could be extended naturally to black-box models where only frequency response data is available. For such cases, frequency domain identification techniques (such as Pade approximations) can fit rational approximations to the measured or simulated (if only precompiled EMT solvers are available) frequency responses of devices, with emphasis on accurately capturing dynamics in critical frequency ranges where non-passivity occurs (typically low frequencies for droop-controlled GFM inverters). Once these rational models are obtained, the multiplier synthesis proceeds exactly as described, enabling practical application even when internal device models are unknown or proprietary.

Our parametric, optimisation-based framework systematically searches dynamic multipliers that certify stability, generalising classical (often conservative) passivity tests and moving beyond ad-hoc piecewise heuristics \eqref{eq:piecewise}.
While the paper focuses on droop-controlled GFM converters, the same methodology can be extended to other converter types and grid components. The key advantage of our approach is achieving these expanded stability guarantees without requiring any hardware or firmware modifications to existing converter implementations—enabling greater flexibility in grid operation while maintaining the PnP property that is essential for scalable deployment.

\section{Case Studies and Results}\label{sec:CaseStudies}

\begin{figure}
  \centering
  \begin{subfigure}[b]{0.8\columnwidth}
    \centering
    \includegraphics[width=0.85\columnwidth]{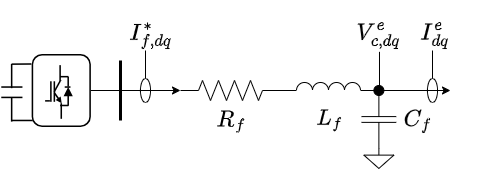}
    \caption{}
    \label{fig:inverter_circuit}
  \end{subfigure}
  \hfill
  \begin{subfigure}[b]{0.95\columnwidth}
    \centering
    \includegraphics[width=0.95\columnwidth]{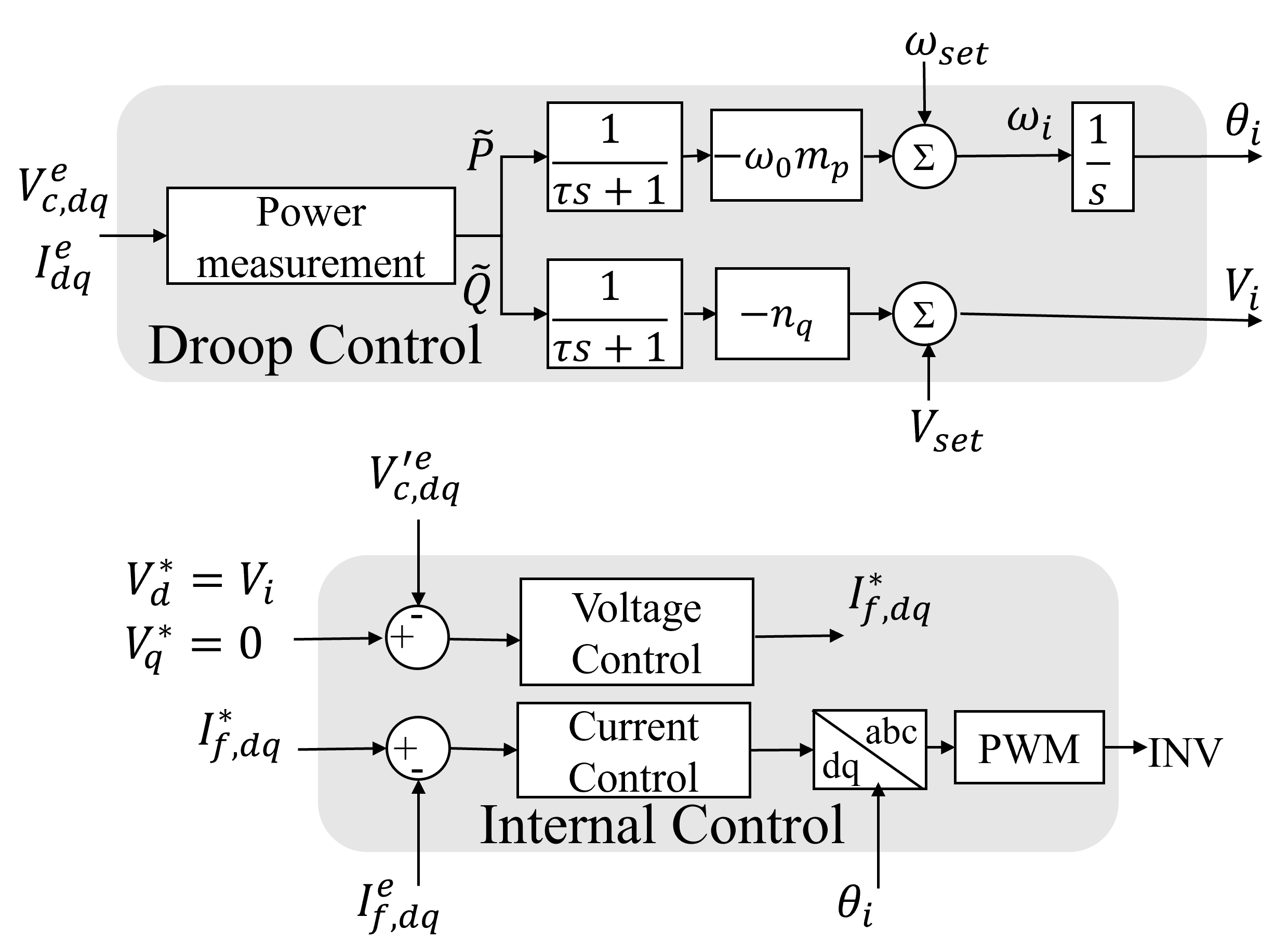}
    \caption{}
    \label{fig:GFM_control}
  \end{subfigure}
  \caption{GFM inverter with LC filter (a) and control architecture (b)}
  \label{fig:inverter_control}
\end{figure}

The GFM control scheme from \cite{pogaku2007modeling} is adopted in this paper shown in Fig. \ref{fig:GFM_control}. The GFM controller comprises three cascaded control loops connected in series: droop, voltage and current. Compared with simplified models that include only droop dynamics, a detailed small-signal model that incorporates the voltage and current loops reproduces the closed-loop dynamics more accurately by considering the coupling effects between control loops.

The droop control adjusts frequency based on active power and voltage based on reactive power:
\begin{align}
\omega &= \omega_0 - m_p(P - P_0),\\
V &= V_0 - n_q(Q - Q_0),
\end{align}
where $\omega_0$ and $V_0$ are nominal frequency and voltage setpoints, $P_0$ and $Q_0$ are the reference active and reactive power outputs, $P$ and $Q$ are the measured powers processed through low-pass filters, and $m_p$ and $n_q$ are the active and reactive power droop coefficients. The voltage computed by the droop law is supplied as the reference to the outer voltage-control loop.
Voltage and current regulation are implemented by cascaded PI controllers, with the outer voltage controller providing the reference to the inner current loop \cite{pogaku2007modeling}. 
The current controller is tuned by a standard pole–zero cancellation procedure so that the controller zero cancels the dominant pole of the $R_f$,$L_f$ filter\cite{yazdani2010voltage}. 
The parameters are given in Table \ref{tab:parameters}.


\begin{figure}
  \centering
  \vspace{-8pt}
  \includegraphics[width=0.8\columnwidth]{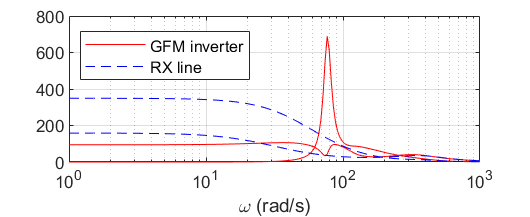}
  \caption{Eigenvalues of~$S_{inv}(j\omega)$ and $S_{line}(j\omega)$ for the tuned multiplier~$m(j\omega)$.}
  \label{fig:eigs}
  \vspace{-8pt}
\end{figure}

\subsection{Two-bus case}
\begin{figure}
  \centering
  \includegraphics[width=0.8\columnwidth]{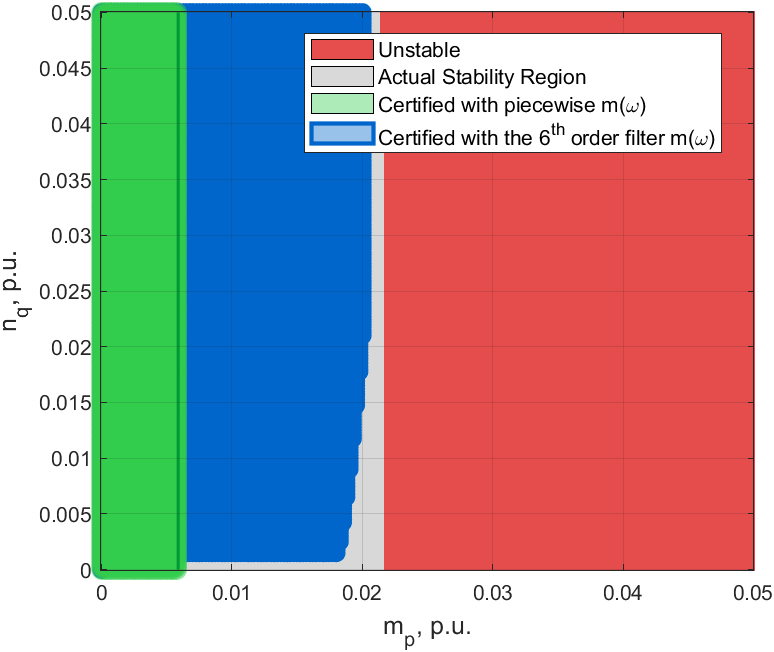}
  \caption{Certified stability region in droop space for the two-bus system.}
  \label{fig:stability_two_bus}
\end{figure}

We first use a representative set-up with parameters given in Table \ref{table:two-bus}. By finding $m(s)$ suitable for both inverter admittance $Y_{\text{inv}}$ and line admittance $Y_{\text{line}}$, this example illustrates the conservativeness of the method. First, we tune a sixth-order $m(s)$ for the given $Y_{\text{inv}}$ with 1\% active and 1\% reactive droop. As shown in Fig.~\ref{fig:eigs}, we successfully certify stability with these droop parameters and line types (actually, any lines with the same $X/R$ ratio are certified by this procedure, as with fixed $X/R$ the worst case is the connection of the inverter with an infinite bus, since any addition of line will deteriorate the stability ~\cite{gorbunov2021identification}). Second, we check numerically the certified region in droop space within practical values (1-5\% for both real and reactive droops) we can certify using this filter, as shown in Fig.~\ref{fig:stability_two_bus} in blue. We also overlay the stability region given by the original heuristic matrices from \cite{vorobev2019decentralized} in green—one can see that the region certified by the tuned sixth-order filter $m(s)$ covers most of the actual stability region in practical range, while the original matrices in green can cover only a portion of up to 0.6\% in active droop $m_p$.

\begin{table}[t]
  \centering
  \caption{System Parameters}
  \label{table:two-bus}
  \begin{tabular}{|c|c|c|}
    \hline
    Parameter & Description & Value  \\
    \hline
    $V_b$ & Base voltage & 1 p.u. \\
    $S_b$ & Base power & 600 MVA \\
    $\omega_0$ & Nominal frequency & $2\pi\times50$ rad/s \\
    $\omega_c$ & Power filter cutoff frequency & 125.6637 rad/s \\
    $K_{pv}$ & Voltage controller proportional gain & 1.7778 p.u. \\
    $K_{iv}$ & Voltage controller integral gain & 0.0 p.u. \\
    $K_{ic}$ & Current controller time constant & 3.1416e+03 p.u. \\
    $C_f$ & Filter capacitance & 1.7778 p.u. \\
    $m_p$ & Active power droop gain & 1\% \\
    $n_q$ & Reactive power droop gain & 1\% \\
    $R_c$ & Coupling impedance & 0.01 p.u. \\
    $X_c$ & Coupling impedance & 0.015 p.u. \\
    \hline
  \end{tabular}
  
  \label{tab:parameters}
\end{table}

\subsection{IEEE 39-bus system}

To demonstrate the practical applicability of our PnP stability certificates, we performed extensive testing on the modified IEEE 39-bus system. The synchronous generators were replaced by ten GFM inverters to mimic an IBR-dominated power system. All GFM inverters are rated at 600 MVA, except the units connected to buses 31 and 39, which are rated at 1000 MVA. All inverters were configured with parameters from Table \ref{tab:parameters}, specifically using droop gains within our certified stability region (1\% active and 1\% reactive droop).

\begin{figure}
  \centering
  \includegraphics[width=0.85\columnwidth]{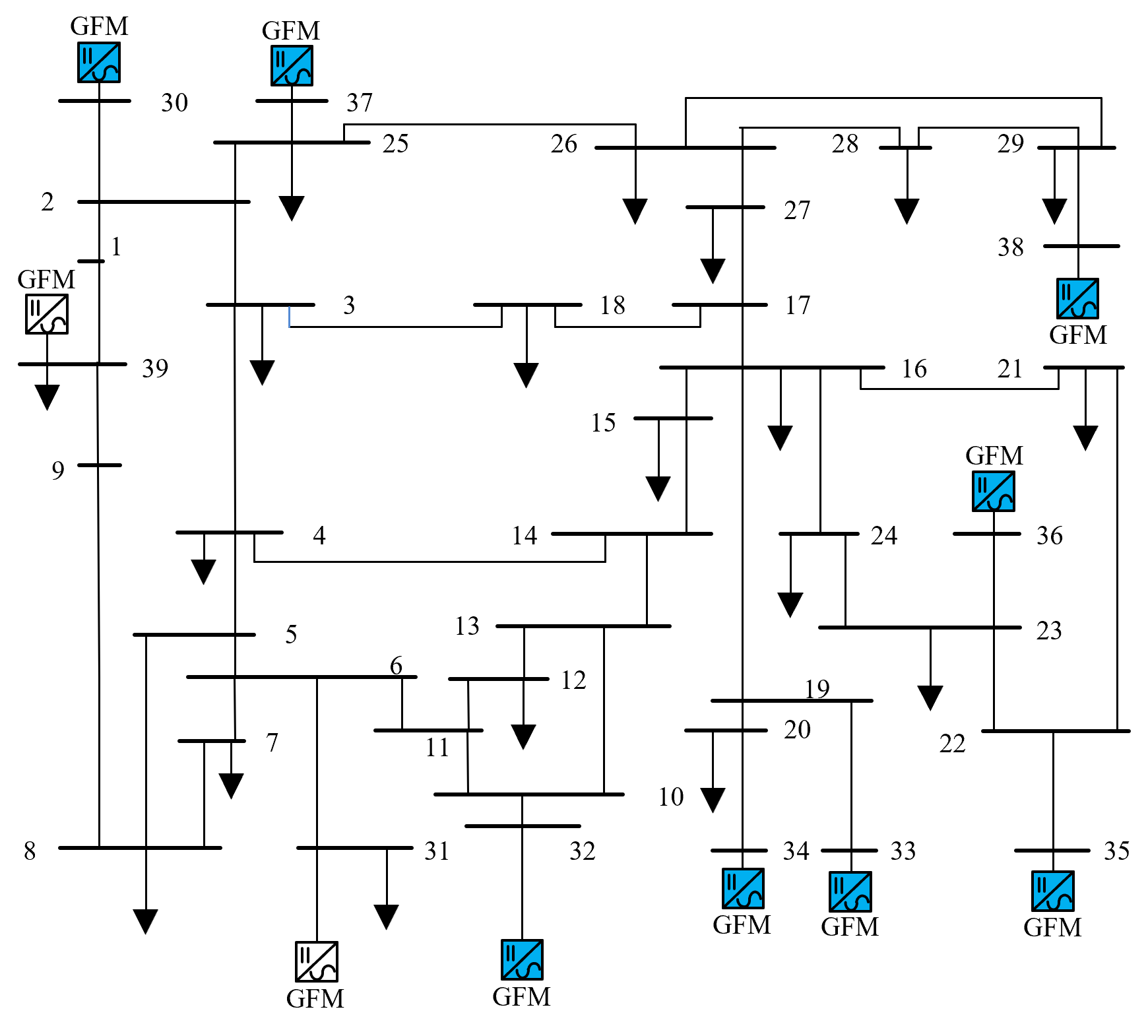}
  \caption{A modified IEEE 39-bus system with 8 GFM converters (blue elements) randomly allocated.}
  \label{fig:ieee39_system}
\end{figure}

\begin{figure}
  \centering
  \includegraphics[width=0.8\columnwidth]{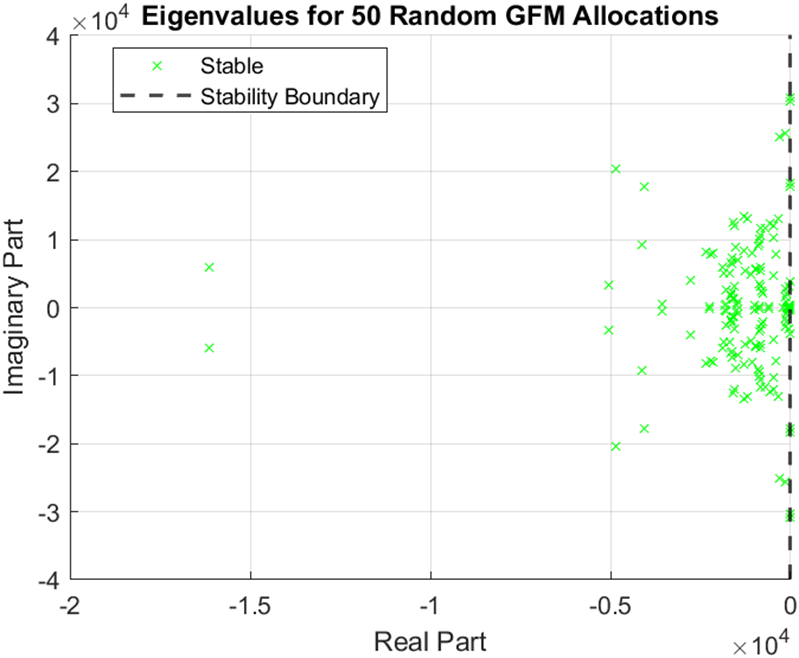}
  \caption{Eigenvalues of the full 39-bus system with 8 GFM converters. 50 random allocations of GFM converters are tested, and all are stable.}
  \label{fig:ieee39_eig}
\end{figure}

The key test of plug-and-play capability is topology independence - stability should be maintained regardless of where compliant devices are connected. To verify this, we conducted 50 trials with random inverter allocations throughout the grid, as one of such realisations is illustrated in Fig. \ref{fig:ieee39_system}. For each allocation, we performed small-signal stability analysis by computing the complete system eigenvalues. As shown in Fig. \ref{fig:ieee39_eig}, all configurations remained stable with eigenvalues consistently in the left half-plane, confirming that our method successfully certifies plug-and-play stability without requiring global system knowledge or coordination. This validates our theoretical finding that decentralised passivity tests with the tuned multiplier guarantee global stability regardless of system topology.

\section{Discussion and Further Research}\label{sec:Conclusion}

This paper develops dynamic-multiplier certificates that extend PnP passivity tests to realistic IBR controllers without requiring firmware changes or detailed network models. We propose an automatic synthesis pipeline using parameterised state-space filters with optimisation-based tuning that systematically derives PnP stability certificates while avoiding excess conservatism. Within this pipeline, we employ a homotopy-free verification that reduces pathwise checks to endpoint conditions, streamlining the search for feasible multipliers. Case studies validate our approach under realistic conditions. In the two-bus case, comparison between certified and numerically stable regions shows only modest conservatism over practical droop gain ranges. In the IEEE-39 system, eigenvalue analyses for randomised inverter placements confirm consistent stability across all certified configurations, demonstrating that our certificates remain valid regardless of device placement or network topology.

The proposed approach has several limitations subject to further investigation:
\begin{itemize}
\item Certification relies on linearised models and guarantees small-signal stability; robustness to large disturbances calls for complementary nonlinear analyses.
\item With incomplete topology/parameter data, worst-case interconnections induce conservatism. Incorporating known bounds while preserving decentralised verification could relax certificates.
\item Algorithm~1 minimises the $\mathcal{H}_\infty$ objective in \eqref{eq:minimax} (a passivity-margin proxy). Higher-order multipliers $m(s)$ typically reduce conservatism, yet the objective does not explicitly maximise the certified parameter region; for the droop-controlled GFM studied here, the optimum attains $1$ (zero margin), resulting in a feasible $m(s)$ rather than one that enlarges the stability region.
\item As a sufficient (not necessary) test, some truly stable designs may yield empty certificates, and a single adverse or unmodelled element can invalidate an otherwise feasible design.
\end{itemize}

Nevertheless, the framework is readily extensible across power-electronic and electromechanical interfaces—including grid-following converters and synchronous machines—suggesting a unified route to stability certification in mixed fleets. Future work will automate multiplier-order selection, exploit partial network knowledge to trade off coverage and conservatism, broaden device classes, and validate performance in high-fidelity real-time simulations.


\bibliographystyle{IEEEtran}
\bibliography{bibtex/bib/bibliography}
\end{document}